\documentclass{article}
\usepackage{array}
\usepackage{color}
\usepackage{graphicx}
\usepackage{authblk}

\begin{document}
\title{Complex versus Complicated Systems Biology, Universality versus Detailed Modelling }
\author{Kunihiko Kaneko}
\affil{Niels Bohr Institute, University of Copenhagen, Blegdmsvej 17, 2100 Copenhagen, Denmark, \\
Universal Biology Institute, University of Tokyo, 3-8-1 Komaba, Tokyo 153-8902, Japan
}
\maketitle
\begin{abstract}Biological systems are generally complicated and/or complex. In the former approach, one sets up a model with a large number of parameters to describe the system in detail. The latter approach focuses on understanding the universal aspects of biological systems. In this case, an appropriate simple model represents a universality class. The extraction of universal properties is supported by evolutionary robustness and the reduction of dimensionality in high-dimensional states. Integrating the data-driven omics approach with the universality approach is an important step in systems biology.
\end{abstract}


\section{Complex versus complicated system}

Biological systems are generally complex and/or complicated. They consist of diverse components and interactions. Here we need to note the difference between a complex system and a complicated system\cite{KK,Oono}. Both are far from simple, but complex systems have some kind of intricate internal order, whereas complicated systems do not have such organized structures, and their description requires many details of their diverse components. No explicit order can be extracted.

Presumably, a biological system has aspects of both complex and complicated systems. Which aspect one should focus on may depend on the aim of the research. If we want to understand life or capture the universal properties that biological behaviors must satisfy, we need to study life as a complex system. However, if we want to predict specific behaviors or design medicines for particular cases, we need to explore the details of complicated biological systems (see Table I).

Now, is it possible to understand such universal behavior in complex biological systems\cite{UB}?
Recall that studies of complex systems have been developed in physics over decades.
In physics, a system is sometimes regarded as a “complex system” when a simple model, described by a few rules (equations) or a few degrees of freedom, can generate behaviors that are not simply expected from those rules. For instance, chaotic dynamics, represented by a set of equations with a few variables, can generate unpredictable behavior whose description often requires stochasticity or complex symbol sequences.
Simple interactions among many identical elements, such as particles or spins, can generate collective behavior or intricate order, as extensively studied in statistical physics of interacting elements. In physics, “complexity from simple rules” is sometimes a motto.

Biological complex systems, however, seem to have different aspects from those studied in physics.
They are not generally described by a few variables or rules. In contrast to statistical physics, which treats a system with a large number of identical elements and simple interactions, biological systems involve a huge number of different components and diverse interactions. Cells generally contain at least a few thousand different proteins, RNAs, and other metabolites, whereas the number of each molecule is not necessarily extremely large, much smaller than the Avogadro number in physics. There is a huge number of reactions among the components. From so many components and interactions, the functions are generated through their coordination, which are often simple, such as responses to signals or adaptation to a given external environment. Indeed, such relatively simple functions could be generated with much simpler systems with few components.

For instance, in adaptation, when an external change is applied, some variables change to compensate for the perturbation, so that other variables return close to the original state. When they return fully to their original values after an external input, this is called perfect adaptation\cite{Koshland}, and many models have been proposed\cite{Asakura-Honda,Barkai-Leibler}. Such behavior can be described by a very simple system, for example a two-component reaction system, with one variable for absorbing the external change and another variable returning to the original state\cite{Koshland,UB}, or by an incoherent feedforward gene-expression network with a three-gene motif\cite{Ma,Alon,Briat}.

However, adaptation in biological cells generally involves many components or genes. Their expression levels often show partial adaptation, in which many of them rise (or decline) and then move back only partially toward the original level, leading to a “collective adaptation process”\cite{Braun1,Braun2}. Models with reaction networks or gene-regulation networks also show such collective adaptation involving many components after evolution\cite{CFKK-adap,Inoue}, even though an optimal network motif for adaptation would be designed by much simpler models. Biological systems often adopt generic adaptation without such optimized structure \cite{UB,Braun1,Koganezawa}, as a collective behavior of many different components.

Recall that collective behavior arising from many degrees of freedom is observed in physical complex systems. Statistical physics have been developed to study how such macroscopic order emerges, which is also relevant to study biological complex systems. However, in biological systems these degrees of freedom are provided by many heterogeneous components, rather than by large numbers of identical elements. This property gives biological complexity a character distinct from that typically discussed in physics.

As another example, in morphogenesis, a reaction–diffusion system with just two degrees of freedom is, in principle, sufficient to generate a stripe pattern, as shown in Turing’s pioneering study\cite{Turing}. However, biological development generally involves the expression of many genes within complex network structures\cite{development}. In this case again, many components are used to produce a relatively simple pattern such as a stripe. In this sense, one may say that in biological systems “simple behaviors are generated from complex systems with many different components,” in contrast to typical physical systems.

In general, collective outputs from diverse elements that are ubiquitous in biological systems generate biological functions\cite{UB}. According to such functions, fitness for survival emerges, with which selection for evolution occurs. This evolution then acts to shape the biological system for the next generation. Hence feedback from the outputs to the system itself follows\cite{Wiener}. This evolutionary shaping is essential to biological complex systems, which are not designed externally and often progresses through bricolage\cite{Jacob}. As a result, biological systems may include many degrees of freedom, which seem to be unnecessary but cannot be removed as they also provide robustness to perturbations. We will discuss this issue in the next section. 

\begin{table}
\begin{center}
\begin{tabular}{|c|c|}
\hline
Complex system & Complicated system\\
\hline
Universal & Detailed\\
\hline
Understanding & Description\\
\hline
Human& AI\\
\hline
Simpler but not too simple & Detailed realistic models\\
\hline
\end{tabular}
\caption{Two approaches to systems biology: Complex versus Complicated}
\end{center}
\end{table}

\section{Dimensional reduction}

\begin{figure}[htbp]
\centering
\includegraphics[height=7cm]{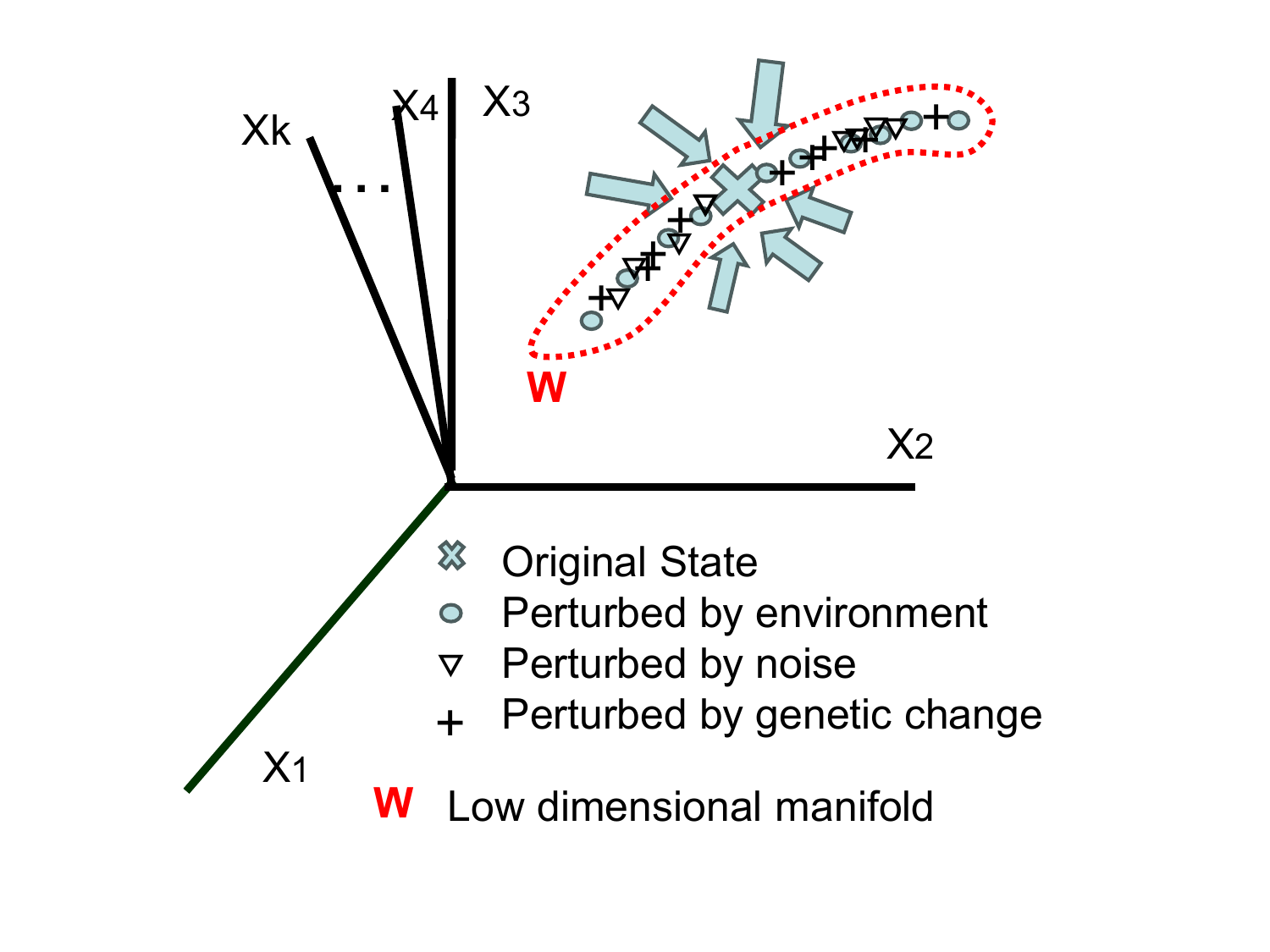}
\caption{Evolutionary dimensional reduction. Even though biological states are potentially quite high-dimensional, most changes induced by noise, genetic changes, and environmental changes are constrained along a common low-dimensional space (manifold), if we focus on a robust state achieved by evolution. The low-dimensional structure as in this figure was observed in a reaction network model of cells after evolution\cite{CFKK-PRE}. The adopted model involves thousand components, and thus the dimension of the system is thousand, but the low-dimensional structure as in the figure was detected by principal component analysis.} 
\end{figure}

Now, can we understand such complex biological systems with many diverse components?
Of course, it would be extremely difficult to fully understand such a high-dimensional system.
Then, in the omics approach, one often describes such systems by simply enumerating all the data.
Will such enumeration be the only solution?
If that were the case, we would not truly understand the system, since understanding generally requires some reduction of such huge data\cite{UB}.

Here we note that in thermodynamics in physics, a drastic reduction of degrees of freedom is possible by restricting our interest to equilibrium thermodynamic states or to approaches toward them. Although the number of molecules is huge, we can represent macroscopic thermodynamic behavior by only a few degrees of freedom, such as temperature, entropy, and pressure.

Of course, biological systems are not in thermal equilibrium. Hence, one cannot rely on the optimism based on thermodynamics. Still, as with the stability of equilibrium states, macroscopic biological systems gain robustness to a variety of external and internal perturbations, and this robustness is achieved through evolution\cite{Wagner,KK-robust,Ciliberti}.
We may then expect that the robustness of a macroscopic cellular state imposes some constraints on the high-dimensional intracellular reaction dynamics,
even though  many degrees of freedom here are associated with diverse components, rather than the number of molecules.

Recall that biological systems are generally hierarchical: molecules → cells → (multicellular) organisms → ecosystems. Here, higher-level behavior is generated as a collective property of lower-level (“microscopic”) units, whereas microscopic levels can also change plastically through feedback from a higher-level (“macroscopic”) state. For instance, molecular composition (and often polymer structure) depends on the cellular state. At the cell–organism hierarchy, cells differentiate depending on surrounding cells in a tissue, etc. With such macro–micro mutual relationships, they form states that are robust to perturbations, as proposed in the “macro–micro consistency principle”\cite{UB}.

Note that there are many kinds of components at the microscopic level: the number of concentration variables of intracellular chemicals or gene expression levels in a cell exceeds thousands, even for a simple cell. Thus, the microscopic state is represented in a high-dimensional state space.
Now let us  consider a steady state of cells in which cells of  essentially same compositions are reproduced after division. This condition imposes a constraint on phenotypic changes, so that the cell growth rate and the replication rates of molecules are balanced\cite{PRX}. In addition, evolutionary robustness further imposes constraints on the changes of many components across different environmental conditions.

To study such constraints, we simulated a cell model consisting of a large number of components whose concentrations change through catalytic reactions determined by a reaction network\cite{CFKK-PRE}. Stationary cellular states are achieved as attractors as a result of the intracellular reaction dynamics under a given environmental condition, i.e., the concentrations of external resources. Then these model cells are evolved by mutation to the reaction networks and by selection of those with higher cell growth rates under the given environmental condition. Then, the evolved cell is put to a variety of environmental conditions, to examine the changes in the cellular states, i.e., the concentration changes of all components. It was then found that these state changes of the cells under a variety of environmental perturbations are constrained within a low-dimensional space (see Fig.1). Even if the microscopic states involve a large number of variables (such as a huge number of gene expression levels), the possible changes upon perturbations are restricted to a low-dimensional state.

The origin of such dimensional reduction has been explained as follows\cite{CFKK-PRE,KKCF-Macro}:
In general, high-dimensional dynamical systems that generate an adapted phenotype are subject to various perturbations: intracellular reactions are noisy, and environmental conditions also fluctuate. Therefore the fitted phenotypic state should be robust to such perturbations. Thus there should be a strong attraction to the fitted state (an attractor) from most directions in the high-dimensional phenotypic space.
However, if this strong attraction occurred from all directions, evolution toward higher fitness would be difficult, because the state could not be easily changed by genetic mutations that slightly modify the dynamical system. Therefore, along the direction(s) of evolution, phenotypes should remain changeable. Accordingly, dominant changes in phenotypes (for example, concentrations of proteins) will be constrained along a low-dimensional space in which evolution has progressed and will continue to progress (see Fig. 1).

In addition to the cell model with catalytic reaction network, the dimensional reduction has also been confirmed in the evolution of gene-regulatory network models with mutual activation or suppression of gene expression levels.
Besides the model simulations, such dimensional reduction as a result of evolutionary robustness has been observed also experimentally,  as is consistent with the above theoretical explanation\cite{CFKK-PRE,KKCF-Macro}. The robust state is strongly attracted from many directions in the high-dimensional space, whereas along a few-dimensional manifold in the state space, plasticity remains so that evolution can proceed. A constraint for a robust system achieved through evolution leads to phenotypic changes caused by noise, environmental changes, and genetic changes being constrained along a common low-dimensional manifold. This also implies that changes in high-dimensional phenotypes caused by adaptation and evolution are aligned, as has been experimentally demonstrated.
This leads to an evolutionary constraint in genotype–phenotype–fitness mapping (see also \cite{Manruiba}).

The dimensional reduction of high-dimensional phenotypic states has attracted much attention in recent years and has been experimentally verified in bacterial responses to environmental stresses or antibiotics\cite{Furusawa2,Sato-KK2023}, in protein dynamics\cite{Tlusty,Tang-KK}, in development\cite{Mani,Jordan,Uchida,Saito,Rohner}, and in organismal motion behavior\cite{Leibler,Stephens}, and so forth.
Here note that the irrelevance of many parameters in high-dimensional biological systems is also discussed in the so-called sloppy-parameter hypothesis\cite{sloppy}, where computational methods to reduce sloppy parameters have been developed\cite{MBAM,FISR}.

As dimensional reduction is expected as a consequence of evolution, the course of phenotypic evolution is constrained along the adaptive response or fluctuations that occur prior to genetic changes\cite{KK-robust,KKCF-Macro,Murugan}. Such theory can be examined by tracking phenotypic changes throughout evolution\cite{Furusawa2,Furusawa-exp}. 
By using transcriptome analysis of {\sl E. coli} cells, such dimensional reduction has also been observed during the course of laboratory evolution, which also allows prediction of phenotypic evolution\cite{Furusawa1,Furusawa-exp,Sato-KK2023}.
It is also interesting to note that similar dimensional reduction is observed in neural systems\cite{Batista}, where changes due to learning, rather than evolution, are constrained.

The possibility of dimensional reduction and of a macroscopic view of biological complex systems was pioneered by Conrad Waddington\cite{Waddington}. He proposed that the cell differentiation process, in spite of the large number of involved genes, can be represented by a landscape in a one- or few-dimensional space, which supports robustness not only of final phenotypes but also of developmental paths, as proposed in “homeorhesis.” This was originally a conceptual proposal, and how developmental dynamics in high-dimensional phenotypic space achieve homeorhesis remains elusive. Recent dynamical-systems models have suggested that dimensional reduction of developmental paths allows the generation of a Waddington landscape and homeorhesis\cite{Matsushita}.

The evolutionary dimensional reduction discussed here will provide the possibility to "understand" a high-dimensional biological complex system, and to construct an effective model as is discussed in the next section.

\section{Modeling}

\begin{figure}
\centering
\includegraphics[height=7cm]{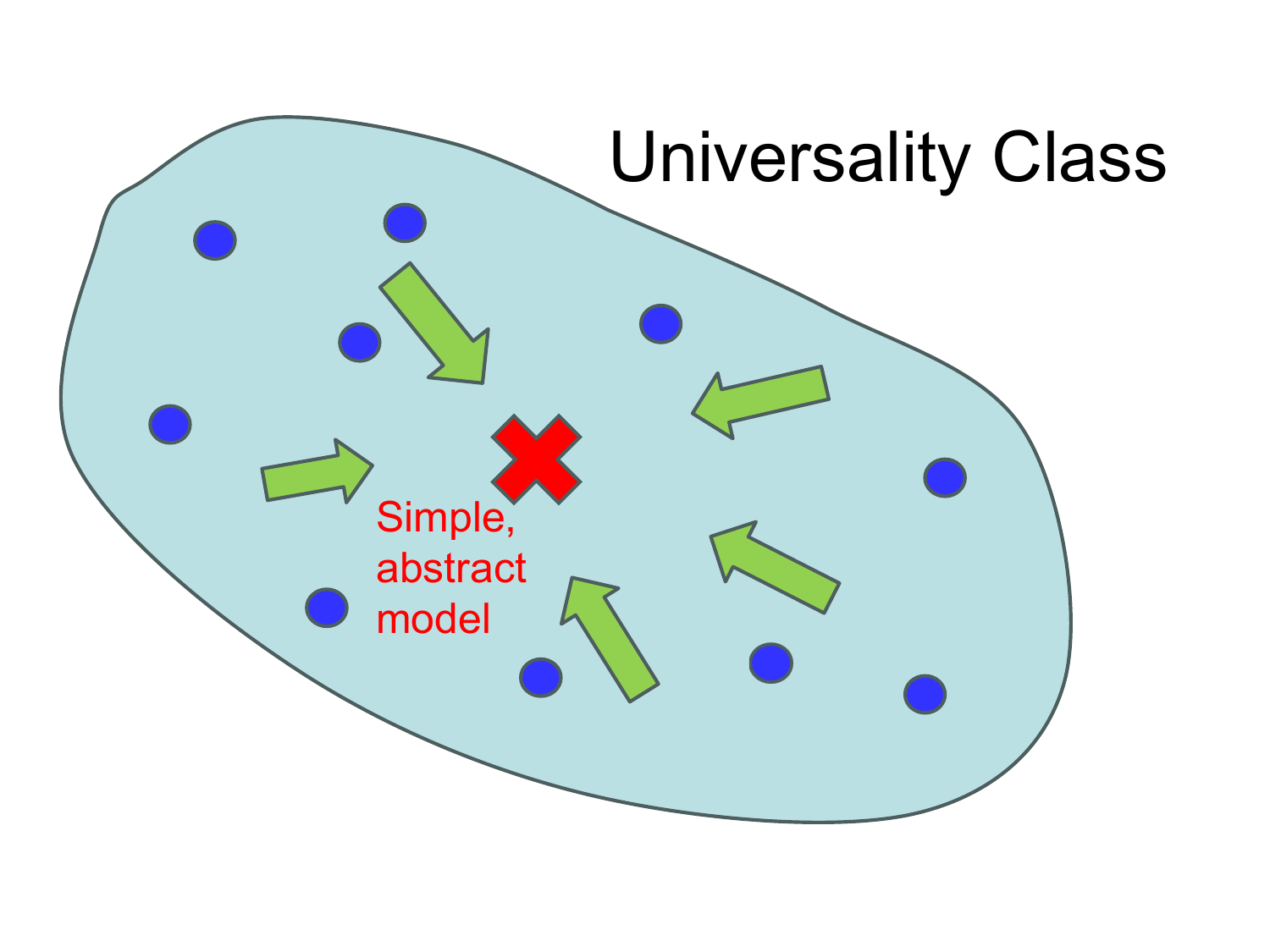}
\caption{Models and universality class.  There are many models that show essentially same behavior, when one focuses on macroscopically universal properties.  Such range of models form a universality class. If some "coarse-graining procedure" to eliminate details is defined, an idealistic simple model can be reached by it.}
\end{figure}

Modeling of biological systems as complex systems in the above sense and as complicated systems requires different strategies. In the latter case, one first obtains detailed information from omics data and sets up a model that includes as much of these details as possible, in order to fit the high-dimensional experimental data. Even though processing such a huge amount of data would be quite difficult for human scientists, it may be possible for AI, which can handle huge amounts of data and generate complicated models by taking advantage of efficient statistical analysis. This seems to be a mainstream direction in current biological studies.

In contrast, in the complex-systems approach, one does not attempt to set up a detailed model that fits the experimental data. Instead, the construction of a model that extracts universal features of the biological system is pursued. Such a model is intended to be simple, but not overly simplified, so that it can capture the essence of biological complexity. 
Often the choice of a relevant model is difficult, since there may be many possible candidate models, and one also needs to decide which details should be extracted or ignored. Therefore, there is no general recipe for selecting the appropriate model.

Now, in statistical physics, there is a concept called a “universality class,” originally developed in the study of critical phenomena. When a phase transition occurs, the scaling behavior of physical quantities around the critical point does not depend on the detailed properties of the system, 
for instance, specific material conditions,
and is determined only by a few characteristics, such as the symmetry and dimension of the system. A range of systems that share a common scaling behavior is termed a universality class. 

How detailed properties become irrelevant to scaling behavior around the critical point is formulated by the renormalization group, in which a coarse-graining procedure to abstract away the details of a system is defined; systems in the same universality class share the same fixed point of this renormalization group\cite{Kadanoff,Goldenfeld}. By investigating a simple model that belongs to a universality class, one can then understand the universal behavior.
So far, however, successful applications of the renormalization group to derive quantitative universality are often limited to the vicinity of critical points. Still, the concept itself is more general and can be relevant for deriving coarse-graining procedures that abstract out system details\cite{Oono}.

Now, can we expect a similar approach for biological systems? In this case, what concerns us is not scaling behavior around a critical point, but universal characteristic behaviors of biological systems with regard to cell growth, adaptation, differentiation, and so forth. In other words, a renormalization-group-like concept for “qualitative behavior” is needed.

Here, each “qualitative universality class” corresponds to 
a class of systems that share a common behavior based on some theoretical concept.
Such general concepts include the hypercycle\cite{Eigen}, the autocatalytic set\cite{Kauffman2}, minority control\cite{KK}, collective adaptation\cite{Inoue}, and attractor selection by noise\cite{UB}. Corresponding to these concepts, simple models have been introduced, such as Kauffman’s Boolean networks for cell differentiation\cite{Kauffman1} and catalytic reaction network models for the power-law abundances in mRNA levels\cite{CFKK-adap}, and so forth.

These models are simplified representations of real biological systems. If some coarse-graining procedure, similar to that in the renormalization group, can be defined, then a wide range of systems may be mapped onto such simple models (see Fig.~2). Although such a “renormalization group” for qualitative universality classes has not yet been established, our intuition obtained from simulations of various models suggests that such a framework may exist.
If this is the case, then by studying simple models that belong to a given class, we can gain intuition about possible universal characteristics and the theoretical concepts corresponding to them, and extract common features of biological systems (see Fig.~2).

In fact, there are classic examples of simple models designed to capture universal behaviors, as mentioned above. From simulations of models in which some setups are modified or additional details are included, we have experienced that such behaviors remain invariant. This gives us some intuition about the existence of qualitative universality classes across different models, in which generic behavior remains invariant against the addition or modification of some details.

Note that such models often involve many degrees of freedom, such as concentrations of chemical components or gene expression levels, which influence each other through a given network or matrix. In adopting such models, “random” networks were sometimes used in earlier studies\cite{Kauffman1,Kauffman2,CFKK-adap}. However, biological networks are not random in general; they are the result of evolution. 
To gain intuition about biological universality classes, then, one therefore needs to consider possible consequences of evolution. For it, it will be relevant to integrate the dynamical-systems approach with evolution to select dynamical-systems according to the output of the dynamical systems: for instance, the networks that govern the dynamics are evolved through mutation and selection depending on the phenotypes generated by the dynamics. As mentioned in the last section, dimensional reduction of phenotypes evolves accordingly, which suggests that behavior after evolution is constrained onto a common low-dimensional manifold, independent of the details of the models.
The evolutionary dimensional reduction abstracts out the details of models, in a way similar to the coarse-graining in the renormalization group.

In contrast, a huge amount of data has been accumulated experimentally in recent years. As already mentioned, there is a general trend to construct complicated models with many degrees of freedom from such data\cite{dl1,dl2}, with the aid of machine-learning tools, even though there could be limitations in such an approach, considering the finiteness of available data as well as noise and errors in the data\cite{Limit}. In addition, even if such models can reproduce or predict experimental results as a black-box input–output relationship, we as human beings may not understand why these outcomes occur, nor can we gain insight into what life is.

For setting up such a detailed model, however, there remains an issue of robustness. The system behavior relevant to living states needs to be robust to changes in parameters. Otherwise, even slight changes would cause the system not to function or to malfunction. One therefore needs to distinguish relevant from irrelevant parameters or variables. 
At this point, integration of the “universality approach” with detailed models derived from omics data may be required. In fact, there is a growing direction to integrate data-driven and universality approaches. For instance, by analyzing huge datasets from omics studies and extracting low-dimensional structures, one can abstract away irrelevant variables and obtain some universal properties.

For instance, Kamei et al.\cite{Kamei} adopted Raman spectroscopy to measure intracellular components globally. Raman spectroscopy does not provide a direct one-to-one correspondence with standard transcriptome or proteome data. Still, if dimensional reduction works, then through the low-dimensional structure, different omics measurements can be related to each other. 
In fact, they demonstrated that the changes in concentrations of most intracellular components are restricted to a low-dimensional manifold. Changes in most components are aligned so as to maintain consistent relationships with each other, which can be termed a homeostatic core\cite{UB,Kamei}. Here omics data are used, not to extract detailed information, but to demonstrate that most components are aligned. Once such a low-dimensional structure is identified, most of the data are no longer necessary. 
In this sense, it is somewhat ironic use of omics: omics data are needed to show that most of the huge data are unnecessary and can be eliminated to obtain universal behaviors. On the other hand, there are a few genes that exhibit condition-specific responses, and identifying them requires analysis of high-dimensional data. Exploration of possible molecular origins for dimensional reduction\cite{mol} will also be important.

Now, modelling that integrates omics data with dimensional reduction will be essential for extracting universal characteristics. Here, the reduction is not introduced merely for our convenience in data analysis. As already discussed, both experiments and simulations have confirmed that such dimensional reduction emerges as a result of evolution, to gain robustness to perturbations and plasticity for the adaptation. In fact, there have already been attempts to integrate omics data analysis with such evolutionary dimensional reduction. These include appropriately reduced models derived from metabolic data\cite{Himeoka} and from developmental systems\cite{Francois}. Thus, in addition to further confirmation of evolutionary dimensional reduction based on omics data\cite{Maeda}, future modeling studies that take advantage of evolutionary dimensional reduction will be needed, in order to extract essential structures from high-dimensional biological data.

This research is supported by and the Novo Nordisk Foundation (NNF21OC0065542). The author would like to thank Chikara Furusawa, Tetsuhiro S. Hatakeyama, Kenichiro F Kamei, Yuich Wakamoto, and James Sharpe for useful discussions. He would be also grateful to anonymous referees for illuminating comments

\end{document}